\documentclass[%
 reprint,
 amsmath,amssymb,
 aps,
]{revtex4-2}
\usepackage[pdftex, pdftitle={Article}, pdfauthor={Author}, bookmarks=true, colorlinks, linkcolor=blue,
urlcolor=blue, citecolor=blue, plainpages=false, pdfpagelabels,
final, breaklinks=true]{hyperref}
\usepackage{graphicx}
\usepackage{dcolumn}
\usepackage{bm}

\begin{document}

\title{Nonreciprocal Multi-mode and Indirect Couplings in Cavity Magnonics}

\author{Chi Zhang,$^1$ Chenglong Jia,$^1$ Yongzhang Shi,$^1$ Changjun Jiang,$^1$ Desheng Xue$^1$}
\author{C. K. Ong $^{1,2}$}
\author{Guozhi Chai$^{1}$}
    \email[Correspondence email address: ]{chaigzh@lzu.edu.cn}
\affiliation{$^1$Key Laboratory for Magnetism and Magnetic Materials of the Ministry of Education, Lanzhou University, Lanzhou, 730000, People's Republic of China. \\$^2$Department of Physics, Xiamen University Malaysia, Jalan Sunsuria, Bandar Sunsuria, 43900, Sepang, Selangor, Malaysia.}

\date{\today} 

\begin{abstract}
We investigate the magnon-photon couplings by employing a small magnet within an irregular resonant cavity, which leads to a desirable nonreciprocity with a big isolation ratio. Moreover, the higher-order couplings between the spin wave modes with the polarized photon modes also exhibit the nonreciprocity. These couplings between polarized photon and spin waves could be regarded as a multi-mode hybridization mediated by cavity photons. We also derive a coupling matrix to predict the characteristics of this kind of indirect coupling. The existence of the indirect couplings broaden the field range of the nonreciprocity of the system. The achieved nonreciprocal multi-mode magnon-photon couplings in a single system offer a feasible method to improve the signal transmission quality.
\end{abstract}

\maketitle

\section{Introduction}
 Cavity magnonics, which connect the magnons (the quanta of spin waves \cite{Nat.Phys.11.453}) and photons (the quanta of electromagnetic waves \cite{Nature.118.874}) through magnon-photon coupling, have been developed as one of the central goals in magnon-based cavity quantum electrodynamics \cite{sciadv.1501286,SolidStatePhys.69.47,SolidStatePhys.70.1,SolidStatePhys.71.117}. The magnon-photon coupling is first proposed by Soykal and Flatt\'e in 2010 \cite{PhysRevLett.104.077202,PhysRevB.82.104413}, and then realized in many experiments since 2013 \cite{PhysRevLett.111.127003,PhysRevLett.113.156401}. So far the magnon-photon couplings were found to be pronounced not only with FMR mode but also with the spin wave resonance mode \cite{J.Appl.Phys.117.053910,J.Appl.Phys.119.023905,PhysRevApplied.2.054002,J.Phys.D.50.205003,J.Phys.D.50.205003,PhysRevB.101.014439,PhysRevB.101.014439,Sci.Rep.7.11511}.

 Among the researches of magnon-photon coupling, some experimental results reveal the indirect couplings in cavity magnonics \cite{Appl.Phys.Lett.109.152405,PhysRevLett.118.217201,PhysRevB.97.014419,PhysRevLett.121.203601,PhysRevB.100.094415,PhysRevB.101.064404}. Hyde et al. experimentally observed an indirect coupling between the two microwave cavity mode photons through a yttrium iron garnet (YIG) sphere and the indirect coupling modes have a higher transmission rate than the two uncoupled cavity modes in 2016 \cite{Appl.Phys.Lett.109.152405}. Then, a kind of indirect coupling between two magnons mediated by cavity photons was reported in detail \cite{PhysRevLett.118.217201,PhysRevB.97.014419}. After that, the indirect interaction between a magnon mode and a cavity photon mode mediated by traveling photons was systematically studied  \cite{PhysRevB.100.094415,PhysRevB.101.064404}. These ideas pave a way to investigate two or more unrelated modes in coupling systems. Meanwhile, the nonreciprocal magnon-photon couplings, as well as the nonreciprocal propagation of photons and magnons \cite{PhysRevX.7.041043,PhysRevX.7.031001,PhysRevApplied.7.024028,PhysRevApplied.7.064014,PhysRevApplied.10.047001,PhysRevB.93.054430,PhysRevLett.120.047201,PhysRevB.100.104427}, might also be benefit for applying the magnon-photon coupling in future quantum coherent information processing. Very recently, the nonreciprocal microwave transmission achieved in an open cavity spintronic system with a considerably large isolation ratio and flexible controllability \cite{PhysRevLett.123.127202}. On this basis, the broadband nonreciprocity can also be realized by locally controlling the magnonic radiation \cite{PhysRevApplied.14.014035,PhysRevApplied.13.044039}.

In this work, we designed a system which provided two types of magnon-photon interactions, one is the coupling between FMR mode and cavity-mode photon, the other is multi-mode high-order coupling which is identified as a multi-mode hybridization mediated by cavity photons. Beyond the multi-mode hybridization, we achieved a desirable nonreciprocity in not only the coupling between FMR mode and cavity-mode photon but also other high order indirect couplings. It shed some light on a relationship between FMR mode and spin-wave-mode magnons.

\section{Experiments}
As shown in Fig. 1(a), the experimental system consists of a resonant cavity, which is made up of the copper, and a piece of single crystal yttrium iron garnet (YIG) wafer, which has a low microwave magnetic-loss parameter and high-spin-density \cite{J.Phys.D.43.264002}. The shape of the cavity is designed as a quadrant stadium so that we can break the time-reversal symmetry (TRS) by introducing a small magnetic material into the cavity \cite{PhysRevLett.103.064101,PhysRevE.81.036205}.
The quadrant stadium is combined with a quadrant with radius as 40 mm and a square with side length of 40 mm. The height of the cavity is 5 mm. We define the intersection of axes $x$, $y$ and $z$ as the origin of coordinates with unit as millimeter. Therefore, we can provide the coordinates of the antenna 1, antenna 2 and the center of YIG wafer as (10 mm, 20 mm, 5 mm), (70 mm, 20 mm, 5 mm) and (15 mm, 20 mm, 2.5 mm) respectively. In addition, with a wafer shape, more spin wave modes can be excited by applying magnetic fields \cite{J.Phys.D.50.205003}. The single crystal YIG wafer has a diameter of 5.64 mm, a thickness of 0.48 mm, a saturation magnetization $M_s$ = 1750 G, and a gyromagnetic ratio $\gamma$ = 2.62 MHz/Oe. The Gilbert damping of single crystaline YIG is as low as $\alpha = 3\times10^{-5}$ \cite{Phys.Rep.229.81}. An electromagnet, whose magnetic field can be varied by changing the magnitude of the applied current, is employed to apply a static magnetic field along $z$ direction. Meanwhile, the input and output microwaves were fed and measured by a vector network analyzer (VNA, Agilent E8363B), which was connected with two microwave cables and two antennas (bought from Allwin technology Inc.). In  our experiments, we chose the seventh eigenmode of 10.05 GHz as the cavity mode and the microwave magnetic field distribution is shown in Fig. 1(b). The black circle stands for the YIG wafer. The arrows describe the directions of the microwave magnetic vectors. The strength of microwave magnetic field at the position of YIG wafer is strong that can induce the strong coupling more easily \cite{Nat.Commun.8.1368,PhysRevLett.121.137203,Appl.Phys.Lett.115.022407}. The damping of this mode is described as $\beta = 2.9\times10^{-3}$.

\begin{figure}[htbp]
\centering
\includegraphics[width=8.6 cm]{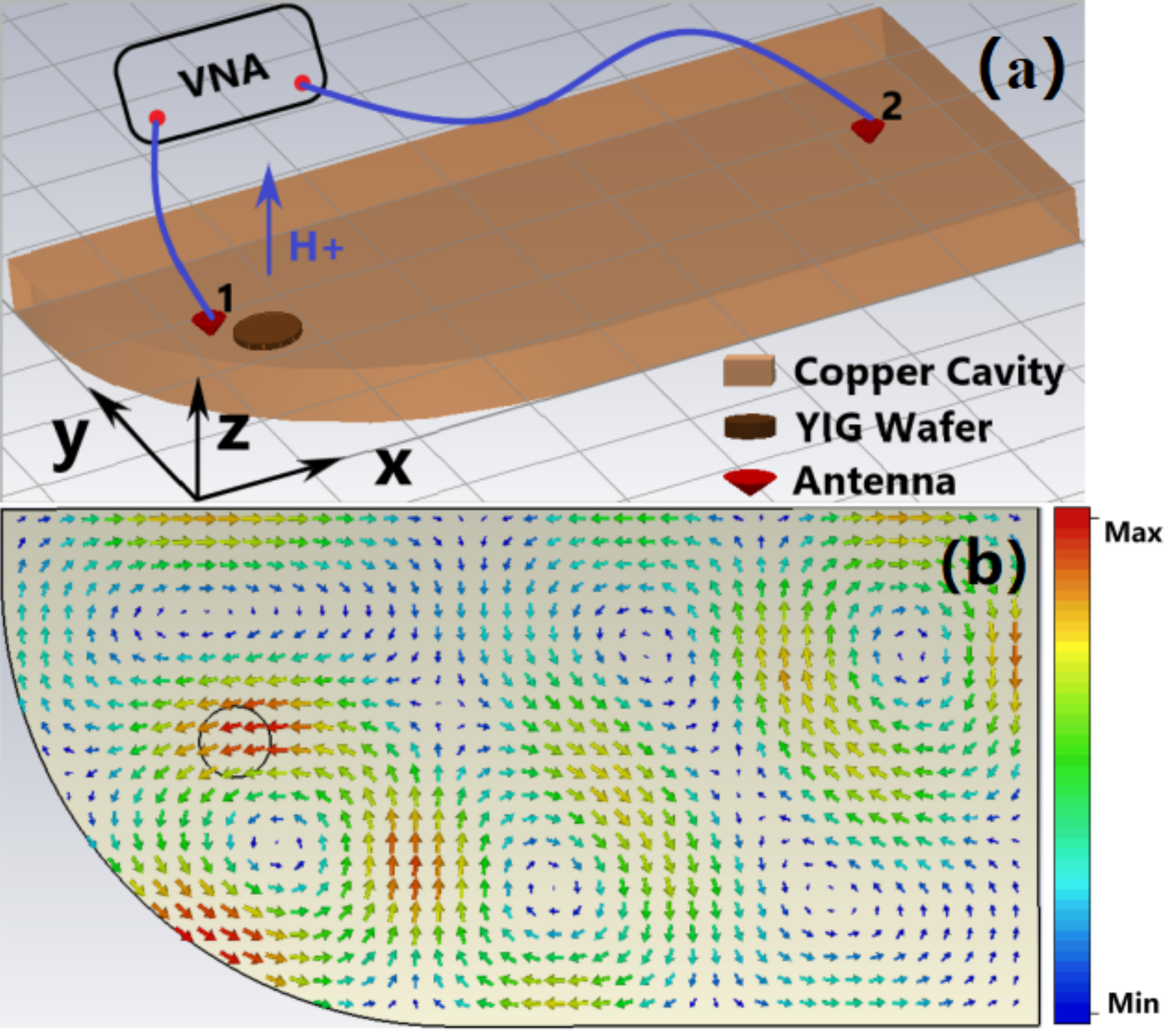}
\caption{(a) Sketch of our experimental setup. A vector network analyzer is employed to provide the microwaves signal into the cavity load with a YIG wafer (YIG-cavity) through two microwave cables with two antennas. An electromagnet is employed to apply a static magnetic field along $z$ direction which is perpendicular to the YIG wafer. (b) Simulation of the microwave magnetic field vectors distribution at 10.05 GHz in the YIG-cavity system. The black circle stands for the YIG wafer. The arrows describe the directions of the microwave magnetic field.}
\label{Fig:1}
\end{figure}

\section{RESULTS AND DISCUSSION}
\subsection{Nonreciprocal couplings}
\begin{figure}[htbp]
\centering
\includegraphics[width=8.6 cm]{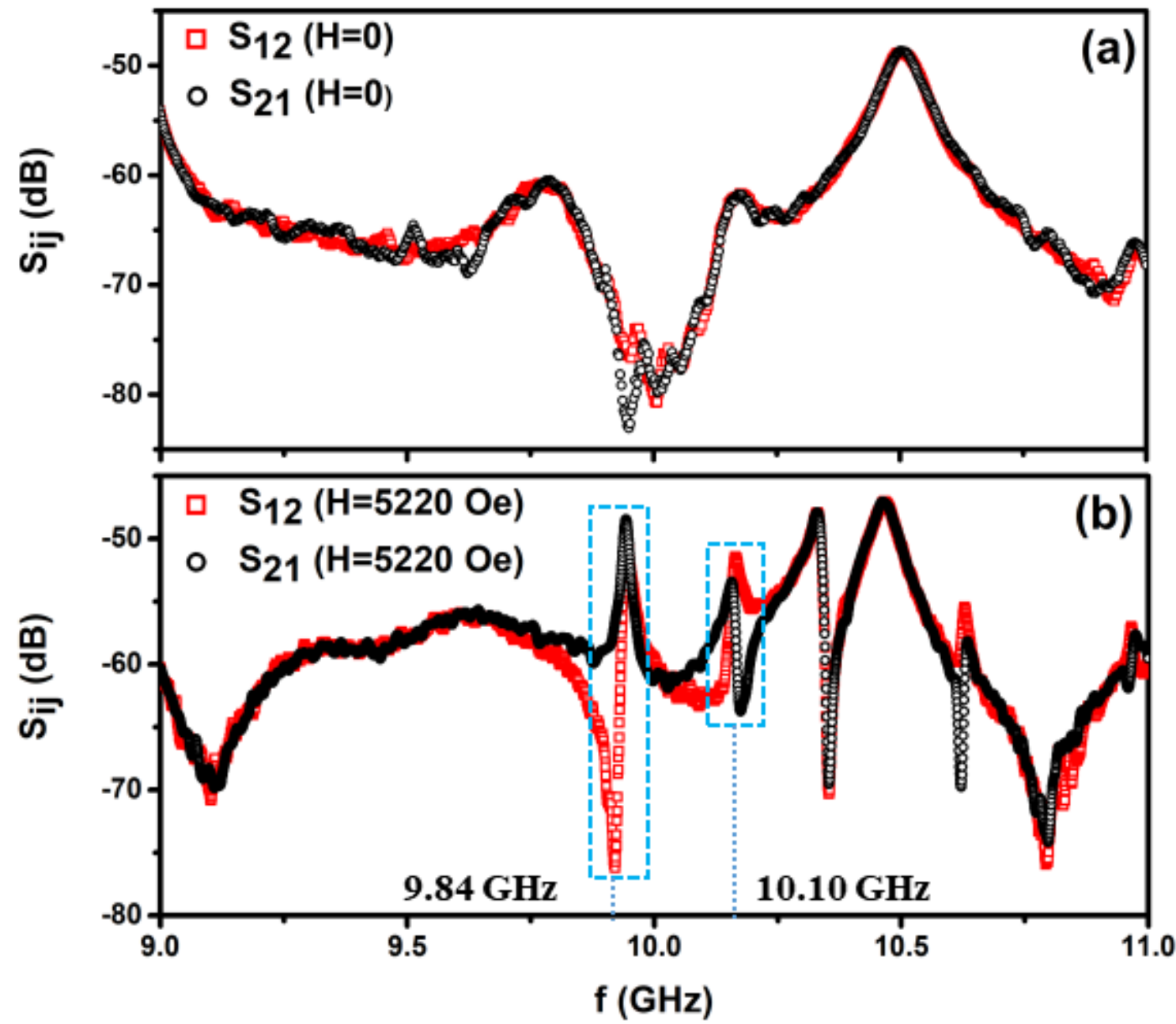}
\caption{Microwave transmission coefficient $S_{21}$ and $S_{12}$ raw data of the YIG-cavity as a function of frequency at the magnetic field of 0 (a) and 5220 Oe (b). Black hollow circles and red hollow squares expresses the microwave transmission coefficient $S_{21}$ and $S_{12}$, respectively. The orientations of the peaks represent the transmission (upward peaks) and loss (downward peaks) of the microwave.}
\label{Fig:2}
\end{figure}

\begin{figure*}[htbp]
\centering
\includegraphics[width=17.2 cm, height=8.6 cm]{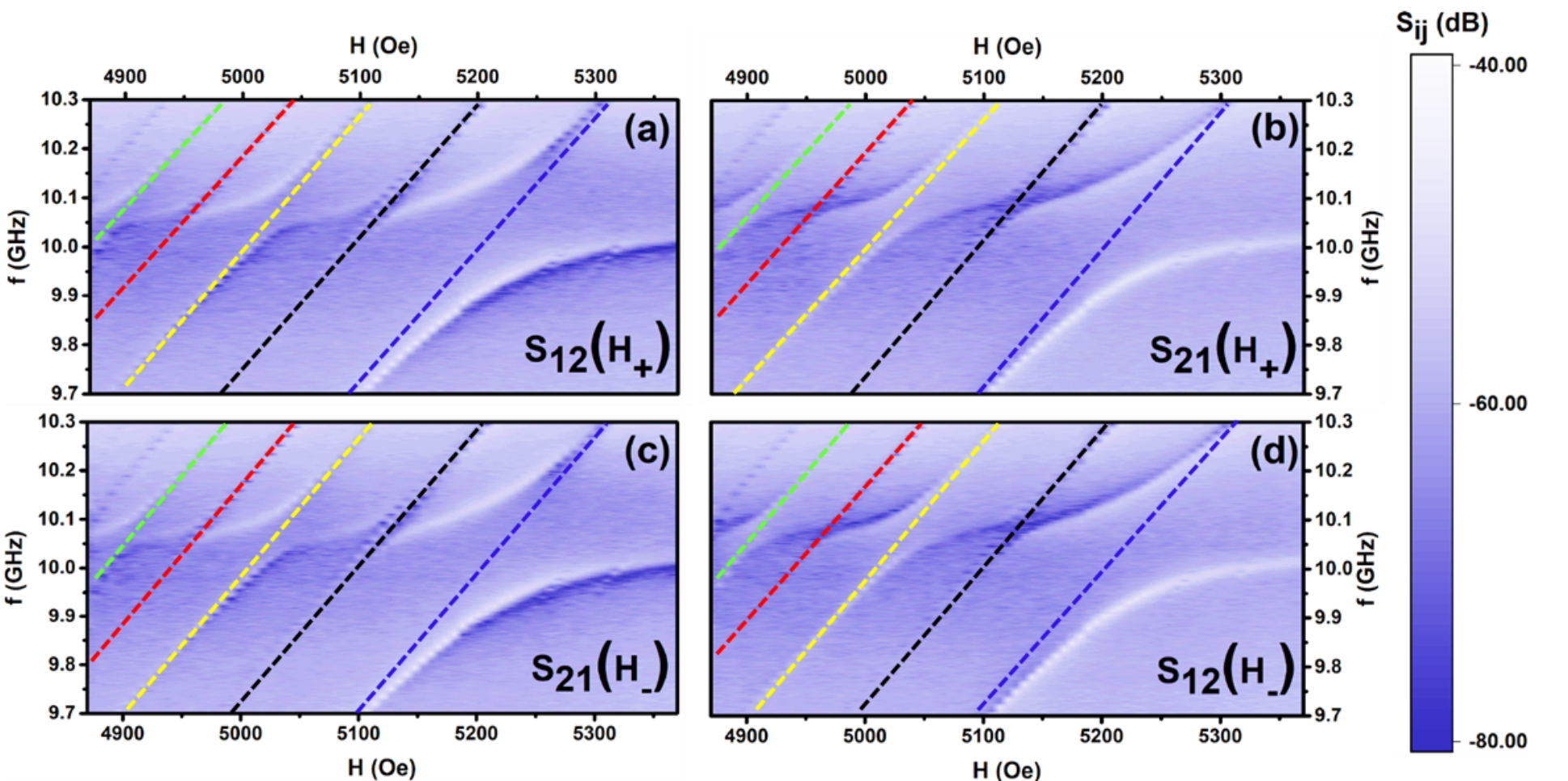}
\caption{The density mapping image of the amplitude of the transmission coefficients through the cavity as a function of frequency and the applied static magnetic field. The deeper colour of image express the larger microwave transmission loss. (a) $S_{12}$ mapping at the magnetic field added along direction +$z$ ($H_+$). (b) $S_{21}$ mapping at the magnetic field added along direction +$z$ ($H_+$). (c) $S_{21}$ mapping at the magnetic field added along direction -$z$ ($H_-$). (d) $S_{12}$ mapping at the magnetic field added along direction -$z$ ($H_-$). Blue dashed line indicates the FMR mode fitting with the Kittel equation [Eq. (1)] and black, yellow, red and green short dashed lines indicate the four FVMSW modes fitting with the FVMSW dispersion relation [Eq. (2)].}
\label{Fig:3}
\end{figure*}
In this paper, we will make use of the scatter parameter $S_{ij}$ (i, j = 1, 2) to characterize the experimental phenomena discussed subsequently. $S_{ij}$ stands for the microwave transmission signal from port $j$ to port $i$. Figure 2(a) shows the microwave transmission coefficients $S_{21}$ raw data (black hollow circles) and $S_{12}$ raw data (red hollow squares) of the cavity loaded with a YIG wafer (YIG-cavity) as a function of frequency without magnetic field. The frequency range in the figure is set from 9 GHz to 11 GHz and the two experimental curves are almost overlapped. Then, we applied static magnetic fields in order to observe the variation between transmission coefficients $S_{21}$ and $S_{12}$. We picked up the most typical results with the applied magnetic field $H$ = 5220 Oe, which are shown in Fig. 2(b). Black hollow circles and red hollow squares express the microwave transmission coefficient $S_{21}$  and $S_{12}$, respectively. Several peaks could be found from two curves. Especially, the orientations of the peaks of the microwave transmission coefficient $S_{21}$ and $S_{12}$ near 9.84 GHz and 10.10 GHz (marked by blue rectangle) are totally opposite. It suggests that this system leads to a nonreciprocal microwave transmission at such a magnetic field. The orientations of the peaks represent the transmission (upward peaks) and loss (downward peaks) of the microwave. Microwaves could not transmit as the microwave transmission coefficients are close to -80 dB.

In order to understand this phenomenon clearly, we swept the applied static magnetic fields with a smaller frequency range between 9.70 GHz to 10.30 GHz. The density mapping image of the magnitude of transmission coefficients are shown in Fig. 3. Figures 3(a) and 3(b) show the density mapping images of the magnitude of $S_{12}$ and $S_{21}$ transmission coefficients as the static field direction is added along +$z$, respectively. Figures 3(c) and 3(d) show the density mapping images of the magnitude of $S_{21}$ and $S_{12}$ transmission coefficients as the static field direction is added along -$z$, correspondingly. The mapping spectra are measured by varying the static magnetic field $H$ from 4870 to 5370 Oe with a step size of 6.8 Oe. The FMR mode $f_{\rm{K}}$ (blue dashed line) is calculated by the Kittel equation \cite{PhysRev.73.155}

\begin{equation}\label{1}
 f_{\rm{K}}=\gamma\sqrt{(H+(N_x-N_z)M_s)(H+(N_y-N_z)M_s)}.
\end{equation}

For the YIG wafer used in this work, the saturation magnetization $M_s$ is 1750 G and the gyromagnetic ratio $\gamma$ is 2.62 MHz/Oe. As the demagnetizing field is not uniform in cylindrical shape magnetized bodies, the demagnetizing factors can not be calculated analytically. We use the experimental results of the demagnetizing factors of the oblate ellipsoid with same dimensional ratio, which is presented in the textbook. $N_x$ and $N_y$ are the demagnetizing factors as 0.07, $N_z$ is the demagnetizing factor as 0.86 \cite{Physics.of.Ferromagnetism}. There are other magnon modes called spin wave modes besides the FMR mode in the system. This kind of spin waves are dominated by the dipolar interaction with large wave lengths. We can confirm they are FVMSW modes because the direction of the applied magnetic field is perpendicular to the YIG wafer \cite{J.Phys.Chem.Solids.19.308,J.Phys.D.43.264002,J.Phys.D.50.205003}. Therefore, we plot the FVMSW mode frequency $f_{\rm{F}}$ versus applied field dependence by the calculation formula \cite{J.Phys.D.43.264002}.

\begin{equation}\label{2}
f_{\rm{F}}=\sqrt{f_{\rm{H}}[f_{\rm{H}}+f_{\rm{M}}(1-(1-e^{-kd_{0}})/kd_{0})]},
\end{equation}
where $f_{\rm{M}}$ = $\gamma M_{s}$, $f_{\rm{H}}$ = $\gamma [H + (N_{x}-N_{z})M_{s}]$, $d_{0}$ is the thickness of the YIG wafer and $k$ is the wave vector. The Bessel function of the first kind of order is introduced to obtain the values of the wave vector. $\mu_{mn}$ stands for the $m$-th eigenvalue of the Bessel function of the first kind of order $n$. In our experiment, wave factor could be defined as $k$ = 2$\mu_{mn}$/$d_{0}$. The value of $\mu_{mn}$ could be found  from the textbook: $\mu_{01}$ = 2.405, $\mu_{11}$ = 3.832, $\mu_{21}$ = 5.136, $\mu_{31}$ = 6.38 \cite{J.Appl.Mech.32.239}. Then we put these values into Eq. (2) and obtained four FVMSW modes as shown in Fig. 3, marked as black, yellow, red and green dashed lines, respectively. Obviously, three anti-crossing phenomena can be found in every mapping image, indicating the normal mode splitting of the hybrid magnon mode and the photon mode. It is worth paying attention that only modes with odd mode number $m$ can induce the couplings. It is because that the modes with even mode number $m$ would cancel coupling strength with the uniform cavity field \cite{J.Appl.Phys.119.023905}. In addition, we find that the couplings shown in the density mapping images of $S_{12}$ and $S_{21}$ transmission coefficients are opposite to each other with the same applied magnetic field, and the TRS is broken completely. The deeper color of image indicates the larger microwave transmission loss. As the direction of the applied magnetic field changed from +$z$ into -$z$, the $S_{21}$ and $S_{12}$ transmission coefficients are exchanged also. It implies a nonreciprocity which is chiral symmetric in this system. In this work, the microwave magnetic field is perpendicular to the applied magnetic field, so that the magnetization intensity shows the gyromagnetism. And we get a tensor permeability $\mu_{ij}$ ($i, j = x, y, z$) described as :
\begingroup
\renewcommand{\arraystretch}{1.5}
\begin{equation}
\mu_{ij} =
\left[
\begin{array}{ccc}
 \mu_{\parallel} & -i\kappa & 0\\
 i\kappa & \mu_{\parallel} & 0 \\
 0 & 0 & \mu_{\perp} \\
\end{array}
\right]
\end{equation}
\endgroup
where $\mu_{\parallel}$ and $\mu_{\perp}$ represent the in-plane permeability and out-of-plane permeability of the YIG wafer, respectively. The off-diagonal terms $\mu_{xy}$ and $\mu_{yx}$ make a significant effect in the vicinity of the resonance field and the gyromagnetism enhances. It results in a nonreciprocal transmission of the electromagnetic wave \cite{PhysRevLett.74.2662}. An interchange of input and output channels change the rotational direction \cite{PhysRevLett.103.064101}. It explains the origin of the nonreciprocity in this work.

\begin{figure}[htbp]
\centering
\includegraphics[width=8.6 cm]{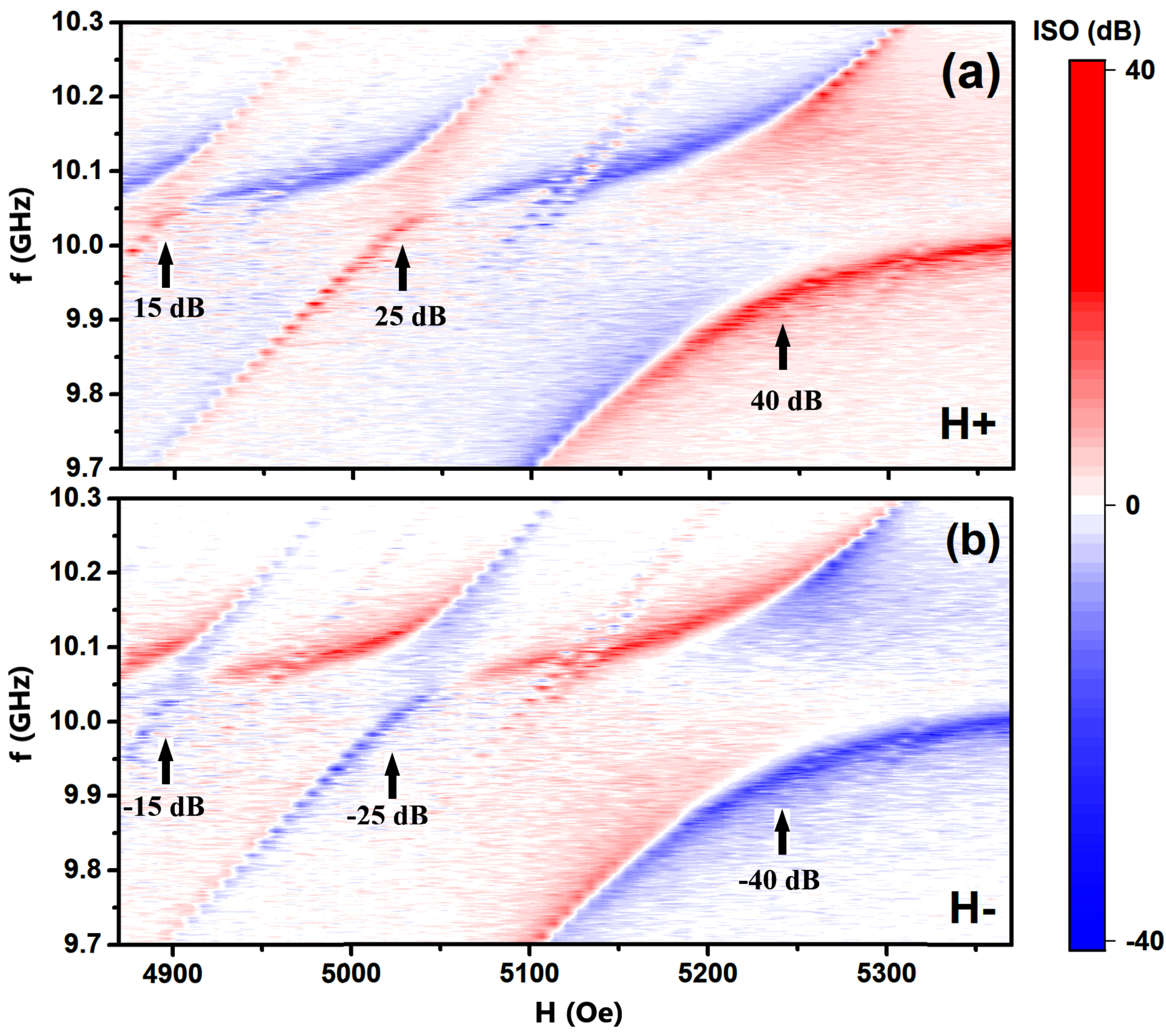}
\caption{ISO of the amplitude of the transmission coefficients $S_{21}$ at the magnetic field along direction +$z$ (a) and -$z$ (b). The largest absolute values of ISO are 40 dB, 25 dB and 15 dB from center coupling to higher order couplings, respectively.}
\label{Fig:4}
\end{figure}

We calculated the isolation ratio (ISO) which is defined as $S_{21}-S_{12}$ so as to express the nonreciprocity clearly. Figures 4(a) and 4(b) represent the ISO of the transmission coefficients at the magnetic field along direction +$z$ and -$z$, respectively. The largest absolute values of ISO are 40 dB, 25 dB and 15 dB from center coupling to higher order couplings, respectively. The ISO of the coupling which induced by the FMR mode with a cavity mode is stronger than other two couplings. It is because the photon modes of these two high-order couplings are dominated by a single linear polarization \cite{PhysRevApplied.13.044039}. It confirms that such a microwave nonreciprocal system with a large isolation ratio could be obtained from an special resonator cavity within a small YIG wafer. The strong reciprocity would occurs in different mode in a much broad range while changing the strength of the magnetic field. The unidirectional invisibility for microwave propagation could be found from the chiral symmetry.

\subsection{Indirect couplings}
In order to comprehend the experiment phenomena more clearly, we need to fit every coupling individually. High order modes and FMR mode are all eigenvalues of the YIG wafer, they can be treated by dividing into single oscillators as eigenvalue of each mode is far away enough. Preliminarily, we judged that our experimental results satisfy this condition from Fig. 3. Then, we fitted the results by the proper equations as shown in Fig. 5. Figure 5(a) shows the couplings in the frequency versus applied static magnetic field with the fitting curves. Blue line stands for the FMR mode $f_{\rm{K}}$ calculated by Eq. (1). Gray and green lines fitting with the FVMSW dispersion relation [Eq. (2)] indicate two FVMSW modes which induce the couplings. The cavity mode frequency $f_{\rm{p}}$ is located at 10.05 GHz (magenta line). The dispersion of the frequencies and the coupling strength $g_1/2\pi$ induced by FMR mode and cavity photon mode can be described as \cite{PhysRevLett.111.127003}
\begin{equation}\label{4}
 f_\pm{_0}=\frac{1}{2}(f_{\rm{p}}+f_{\rm{K}})\pm\frac{1}{2}\sqrt{(f_{\rm{p}}-f_{\rm{K}})^2+(g_1/2\pi) ^2}.
\end{equation}
By fitting the experimental results as shown in Fig. 5 with orange dash lines, $g_1/2\pi$ = 0.23 GHz. Then, we tried to fit the high-order couplings by the traditional two-mode model described as \cite{PhysRevLett.111.127003,J.Phys.D.50.205003}
\begin{equation}\label{5}
 f_\pm{_1}=\frac{1}{2}(f_{\rm{p}}+f_{\rm{F}})\pm\frac{1}{2}\sqrt{(f_{\rm{p}}-f_{\rm{F}})^2+(g_2/2\pi) ^2},
\end{equation}
where $f_\pm{_1}$ are frequencies of the coupled resonances, $g_2/2\pi$ is the coupling strength. The fitting curves are shown in Fig. 5(a). It is clearly that the fitting curves of the high-order couplings do not agree well with the experimental results. It indicates that high order magnon-photon couplings may not be directly interacted by the FVMSW modes with the fixed cavity modes base on the traditional two-mode model. A new coupling model need to be put forward.
As shown in experimental results, different magnon modes can not coupled directly. In this way the two magnon modes might be indirectly coupled through the cavity mode in this system. We study the coupled harmonic oscillator system which Paul et al. gave out \cite{Appl.Phys.Lett.109.152405} and propose out our matrix equation as Eq. (6) to predict the characteristics of indirect coupling between magnon modes via cavity photon mode in this system. Here the diagonal terms are the uncoupled resonance conditions of the two magnon modes and the cavity mode. The off-diagonal terms are the coupling strengths. Two zeros indicate that there is no direct coupling between the two magnon modes. Here the cavity mode is modelled as a mechanical oscillator with amplitude $h_{f}$, and the FMR and FVMSW modes in YIG are described by the mechanical oscillator with amplitude $m_{\rm{K}}$ and $m_{\rm{F}}$, respectively. $\Gamma$ denotes the impedance matching parameter for the cavity mode. $h_{0}$ is the microwave field used to drive resonance in the cavity and is eliminated by normalization in the microwave transmission. This kind of indirect coupling is that after one of the couplings, the energy of the oscillator is changed. Then, the other coupling will occur between the coupled oscillator and other type oscillator. It could be regarded as a multi-mode hybridization mediated by cavity photons rather than a direct coupling between different magnon modes.

\begin{widetext}
\begingroup
\renewcommand*{\arraystretch}{1.5}
\begin{equation}
\left(
\begin{array}{ccc}
 f^2-f_{\rm{K}}^2+i2\alpha f_{\rm{K}}f & g_{1}^2f_{\rm{K}}^2 & 0\\
 g_{1}^2f_{\rm{K}}^2 &f^2-f_{\rm{P}}^2+2i\beta f_{\rm{P}}f & g_{2}^2f_{\rm{F}}^2 \\
 0 & g_{2}^2f_{\rm{F}}^2 & f^2-f_{\rm{F}}^2+2i\alpha f_{\rm{F}}f \\
\end{array}
\right )
\left(
\begin{array}{ccc}
 m_{\rm{k}}\\
 h_{\rm{f}}\\
 m_{\rm{F}}\\
\end{array}
\right )
=
f^2
\left(
\begin{array}{ccc}
 0\\
 \Gamma\\
 0\\
\end{array}
\right )
h_{0}
\end{equation}
\endgroup
\end{widetext}

Comparing with the FMR mode, there is an obvious attenuation with the coupling strength induced by the FVMSW magnon mode. Hence the coupling strength induced by the FVMSW magnon mode does not have a great influence on the cavity photon mode. To further understand the nature of this indirect interaction between the two magnon modes, a coupled harmonic oscillator system is used to model the $f$-$H$ dispersion, damping evolution, and amplitudes during indirect coupling. The frequencies of the coupled resonances $f^{'}_\pm{_i}$ ($i = 1, 2$) and coupling strengths $g_k/2\pi$ ($k = i + 1$) of the couplings between the FVMSW modes and polarized photon modes can be described as
\begin{equation}\label{7}
 f^{'}_\pm{_i}=\frac{1}{2}(f^{'}_{+j}+f_{\rm{F}})\pm\frac{1}{2}\sqrt{(f^{'}_{+j}-f_{\rm{F}})^2+(g_{k}/2\pi) ^2}.
\end{equation}
$j = i - 1$ and $f^{'}_{+0}$ could be regarded as $f_{+0}$ approximately because the coupling strengths induced by the high-order modes have a small effect on FMR mode. The fitting curves are shown in Fig. 5(b). For the first indirect coupling which is induced by the multi-mode hybridization, the olive curve and gray lines are used to describe the polarized cavity photon mode and FVMSW mode, respectively. $f^{'}_\pm{_1}$ are described by the purple curves and $g_2/2\pi$ is the coupling strength with a value of 0.11 GHz. Similarly, the cyan curve and green line stand for the polarized cavity mode and FVMSW mode in second indirect coupling. The frequencies of the coupled resonances $f^{'}_\pm{_2}$ are described by the red curves. $g_3/2\pi$ is the coupling strength with a value of 0.08 GHz. In order to demonstrate whether the eigenvalue of each magnon mode is far away enough, we did some numerical calculation. $\Delta f_{1}$ is introduced to describe the frequency difference of FMR mode (blue line as shown in Fig. 5) and the FVMSW mode (gray line as shown in Fig. 5). Similarly, we define the frequency difference of two FVMSW modes (gray and green lines as shown in Fig. 5) as $\Delta f_{2}$. Then, we calculate the value of $\Delta f_{1}$ and $\Delta f_{2}$ as 0.56 GHz and 0.32 GHz, respectively. Comparing to the strength of the coupling of FMR mode and corresponding spin wave mode with value 0.23 GHz, 0.11 GHz and 0.08 GHz, it is obvious that $\Delta f_{i}$ $>$ $(g_{i}+g_{i+1})/2\pi$ ($i$ = 1, 2), so the eigenvalues of the magnon modes are far away with each other. On this basis, we are able to treat these magnon modes by dividing into single oscillators. In addition, Our FVMSWs are different to another kind of spin wave called perpendicular standing spin waves (PSSWs) with short wavelengths. Cao et al. research that as the thickness of the sample increases, PSSW modes are close to each other gradually and then cannot be treated by dividing into single oscillators \cite{PhysRevB.91.094423}.

\begin{figure}[htbp]
\centering
\includegraphics[width=8.6 cm]{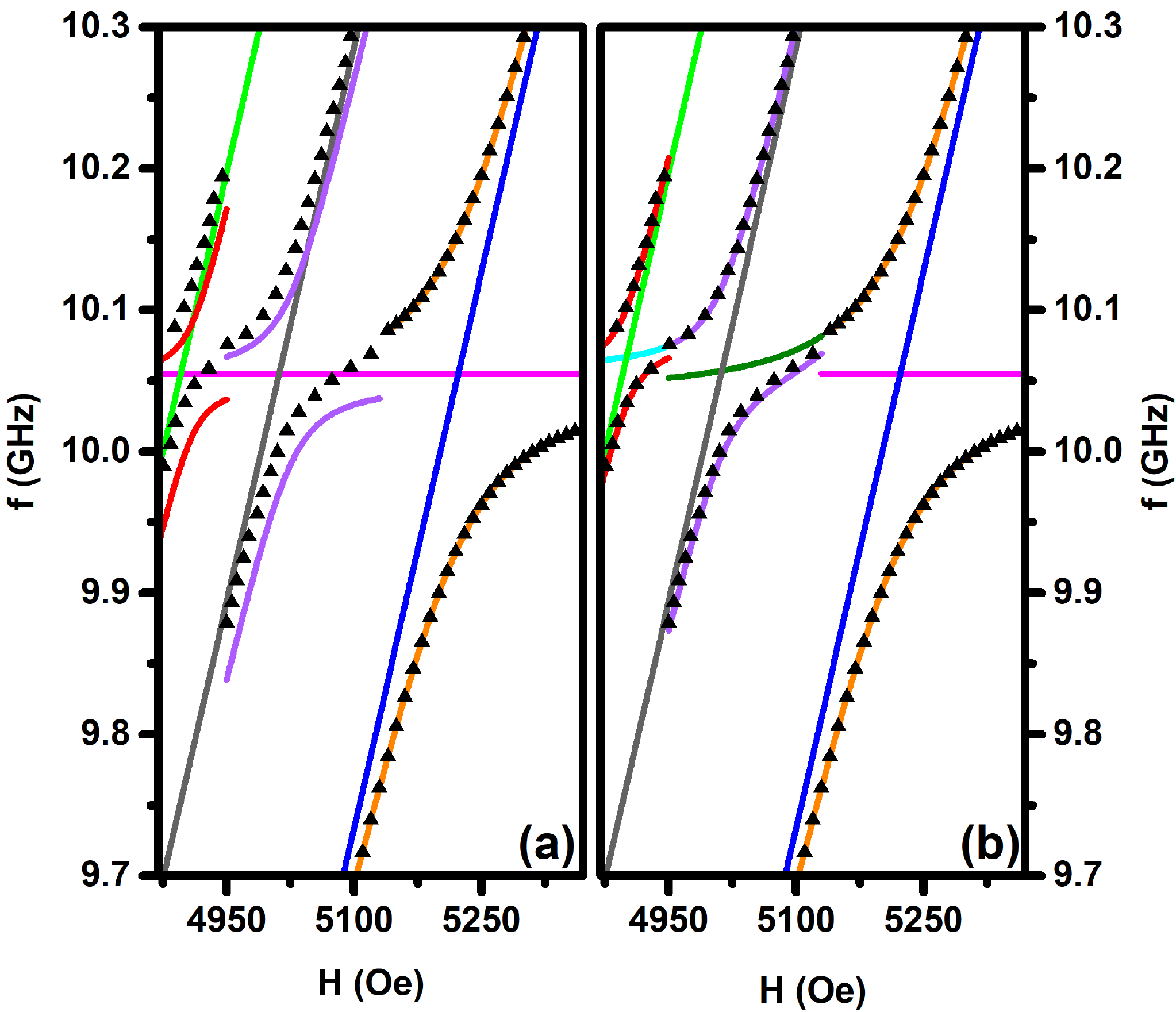}
\caption{The couplings in the frequency versus applied static magnetic field with the fitting curves. The experiment data are described as black triangles. Blue line indicates the FMR mode fitting with the Kittel equation [Eq. (1)]; gray and green lines fitting with Eq. (2) describe two FVMSW modes which induce the couplings; magenta line indicates the cavity mode frequency located at 10.05 GHz; orange curves are fitting with the coupling equation [Eq. (4)]. (a) Purple curves are fitting with the traditional two-mode model described as Eq. (5) and so are the red curves. (b) Olive and cyan curves indicate the polarized photon modes fitting with $f^{'}_{+0}$ and $f^{'}_{+1}$ in Eq. (7), respectively. Purple and red curves are fitting with the multi-mode model described as Eq. (7).}
\label{Fig:5}
\end{figure}

\begin{figure}[htbp]
\centering
\includegraphics[width=8.6 cm]{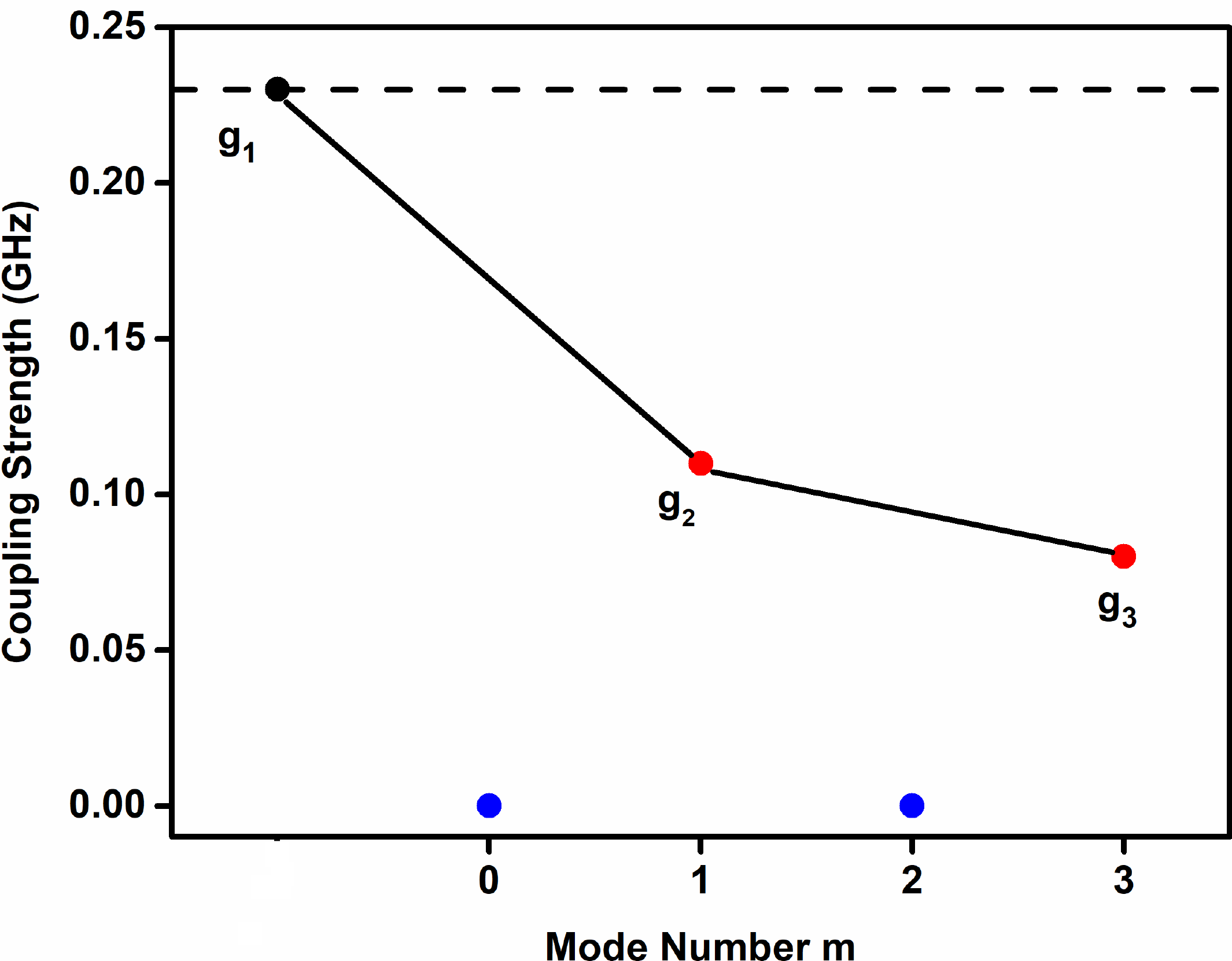}
\caption{Coupling strengths. Dashed line and black circle indicate the strength of the coupling induced by the FMR mode with the cavity mode. Red circles indicate the strengths of the high-order couplings with odd mode number $m$. Blue circles indicate the strengths of the high-order couplings with even mode number $m$.}
\label{Fig:6}
\end{figure}

The coupling strengths changing with increasing mode number are shown in Fig. 6. It shows all of the coupling strengths discussed above. Dashed line and black circle describe the coupling strength of the coupling induced by the FMR mode with the cavity mode. Red and blue circles describe the coupling strengths of the high-order couplings with odd and even mode number $m$, respectively. Besides the results that already been discussed, the relationship between these coupling strengths also arouse our interest and we analyzed it combining the quantum electrodynamic system. The coupling strength $g_{i}$ ($i$ = 1, 2, 3) obeys the $\sqrt{N}$ scaling law, where $N$ is the atom number. Accordingly, the coupling strength in the magnon-photon coupling systems could be regarded as $g_{i} = g_{0i}\sqrt{N}$ ($i$ = 1, 2, 3), where $N$ is the net spin number of the magnet and $g_0$ is the coupling strength of a single Bohr magneton to the cavity \cite{PhysRevLett.113.156401,PhysRevLett.113.083603,PhysRevLett.103.083601,Appl.Phys.Lett.115.022407}.
The net spin density of the pure YIG is $2.1\times 10^{22} \mu_B/{\rm{cm}}^3$, the total Bohr magneton number of our YIG wafer can be calculated as $2.66\times 10^{20}$. As we calculated the single spin coupling strength of FMR-cavity mode, $g_{01}/2\pi$ is calculated to be 15.5 mHz. Comparing with the theoretical value calculated by $g_0/2\pi=\gamma\sqrt{\mu_0\hbar\omega/V}$ = 38 mHz \cite{PhysRevLett.113.083603}, where $\mu_0$ is the permeability of vacuum, the coupling efficiency of our system is 40.8\%. In the same way, we calculated the single spin coupling strength of other higher-order modes as $g_{02}/2\pi$ = 7.4 mHz and $g_{03}/2\pi$ = 5.4 mHz. The total single spin coupling strength of these three couplings is about 28.4 mHz, with the total coupling efficiency as 74.7\%.

\section{SUMMARY}
In summary, we obtained two types of magnon-photon interactions by designing a cavity which can be used to break the TRS easily with a YIG wafer. Not only the usual coupling between FMR mode and cavity mode but also a multi-mode hybridization mediated by cavity photons show an admirable nonreciprocity with a large isolation ratio. Although the single spin coupling strength of the coupling between FMR mode with the cavity mode is about 60\% smaller than the theoretical value, the total single spin coupling strength of these three coupling is gradually approaching to the theoretical value. Compared with the former researches about the nonreciprocal magnon-photon couplings, appearance of such indirect couplings widened the range of the nonreciprocity in a single system, which may be beneficial for quantum unidirectional of processing information and the unidirectional devices which can be used in a wide magnetic field range. It also paves a way to build a relationship between two unrelated magnon modes which can exert a positive influence on the development of cavity magnonics devices.

\section*{Acknowledgements}
The authors would like to thank Z. J. Tay for useful discussions and suggestions.
This work is supported by the National Natural Science Foundation of China (NSFC) (Nos. 51871117), Natural Science Foundation of GanSu Province for Distinguished Young Scholars (No. 20JR10RA649) and the Program for Changjiang Scholars and Innovative Research Team in University (No. IRT-16R35).

\bibliography{manuscript2}
\end{document}